\RequirePackage{fix-cm}
\documentclass{svjour3}                     
\usepackage{graphicx}
\usepackage{mathptmx}      
\usepackage{graphicx}
\usepackage{microtype}
\usepackage[separate-uncertainty=true]{siunitx}
\usepackage{booktabs}
\usepackage{xcolor}

\newcommand{\NO}{(TM\-TTF)$_2$\-NO$_3$}
\newcommand{\ClO}{(TM\-TTF)$_2$\-ClO$_4$}
\newcommand{\TTFX}{(TM\-TTF)$_2X$}

\journalname{Applied Magnetic Resonance}
\begin{document}

\title{
Charge-order phase transition in the quasi one-dimensional organic conductor \NO{}
}


\author{Lena~Nadine~Majer         \and
        Bj\"{o}rn~Miksch          \and
        Guilherme~Gorgen~Lesseux  \and
        Gabriele~Untereiner       \and
        Martin~Dressel
}


\institute{1.~Physikalisches Institut\\
            Universit\"{a}t Stuttgart\\
            Pfaffenwaldring 57\\
            70569 Stuttgart\\
            Germany\\[1mm]
            Corresponding author: Lena Nadine Majer\\
            \email{lena-nadine.majer@pi1.physik.uni-stuttgart.de}\\[1mm]
            Guilherme Gorgen Lesseux\\
            \emph{Current address:} Iowa State University and Ames Laboratory, Ames, IA 50011, USA
}

\date{Received: date / Accepted: date}

\maketitle

\begin{abstract}
Low-dimensional organic conductors show a rich phase diagram, which has, despite all efforts, still some unexploed regions. 
Charge ordered phases present in many compounds of the \TTFX{} family are typically studied with their unique electronic properties in mind.
An influence on the spin arrangement is, however, not expected at first glance. Here, we report temperature and angle dependent electron spin resonance (ESR) measurements on the quasi one-dimensional organic conductor  \NO{}. 
We found that the  \NO{} compound develops a peculiar anisotropy with a doubled periodicity ($ab'$-plane) of the ESR linewidth below about  $T_\text{CO}=\SI{250\pm 10}{K}$. 
This behavior is similar to observations in the related compounds \TTFX{} ($X$ = PF$_6$, SbF$_6$ and AsF$_6$), where it has been attributed to relaxation processes of magnetically inequivalent sites in the charge-ordered state. 
For the structural analogous  \ClO{}, known for the absence of charge order, such angular dependence of the ESR signal is not observed. 
Therefore, our ESR measurements lead us to conclude that a charge-order phase is stabilized in the title compound below $T_\text{CO} \approx \SI{250}{K}$.

\end{abstract}

\clearpage

\section*{Declarations}
\paragraph{Funding}
Financial support by the Deutsche Forschungsgemeinschaft (DFG), grant DR228/52-1, is thankfully acknowledged.

\paragraph{Conflicts of interest}
The authors declare that they have no conflict of interest.

\paragraph{Availability of data and materials}
The data that support the findings of this study are available from the corresponding author, Martin Dressel, upon reasonable request.

\paragraph{Authors' contributions}
All authors contributed to the study conception and design. 
Crystals have been grown by Gabriele Untereiner, ESR measurements and data analysis were performed by Lena Nadine Majer, Bj\"{o}rn Miksch and Guilherme Gorgen Lesseux.
The first draft of the manuscript was written by Lena Nadine Majer and all authors commented on previous versions of the manuscript. 
All authors read and approved the final manuscript.

\clearpage

\section{Introduction}
Low-dimensional organic conductors have drawn attention for decades because they show a rich phase diagram with a variety of interesting ground states \cite{LebedBook}. 
Due the reduced dimensionality of the electronic structure and rather strong electronic correlations, these materials exhibit unusual thermodynamic, transport, optical, and magnetic properties. 
The charge and spin degrees of freedom in the Bechgaard and Fabre salts can be tuned from localized to itinerant by changing the anions as well as external parameters such as temperature or pressure. 
This can lead to various phases in the charge and spin sector, such as charge order, antiferromagnetic phases, spin density wave and even superconductivity \cite{Dressel07,Kohler11,Dressel12c,Rosslhuber18}.
Here we focus on the \TTFX{} family, where TMTTF denotes tetramethyltetrathiafulvalene and $X$ stands for a monovalent anion -- here specifically on $X^-=\text{NO}_3^-$.
Using electron spin resonance (ESR) as local probe of electronic and magnetic properties, we yield information on the phase transitions as a function of temperature.

One-dimensional organic conductors can be described by the Hubbard model. The \TTFX{} compounds are nominally $3/4$-filled systems, dimerization along the chains however leads to a half-filled conduction band.
Hence the compounds are supposed to exhibit metallic behavior, but strong electron-electron interaction can cause localization of the electrons, depending on the competition between electronic correlations modeled by an on-site Coulomb repulsion $U$ and the bandwidth $W$. In addition, most members of the Fabre family
undergo a phase transition to a symmetry broken ground state at low temperatures.
In the present case, one expects that inter-site Coulomb repulsion $V$ drives the systems into a charge-ordered phase;
i.e.\ an alternating arrangement of charge rich and charge poor molecules leads to localization of carriers. Since in the \TTFX{} crystals the stacks are slightly dimerized on stoichiometric grounds,
one does not necessarily expect a doubling of the unit cell \cite{Dressel07}. As continuously stressed by Pouget and collaborators \cite{Pouget96,Pouget12a,Pouget12b,Pouget18}, the interaction with the anions cannot be ruled out and any change in the molecular charge pattern will affect the structural arrangement and vice versa.

The magnetic properties of \TTFX{} salts have been intensely studied by ESR during the last years \cite{Dumm00a,Dumm00b,Nakamura03,Coulon04,Furukawa09,Salameh11,Yasin12,Dressel12a,Coulon15,Dutoit18}.
In the case of centro-symmetric anions, such as PF$_6^-$, SbF$_6^-$ and AsF$_6^-$, Yasin {\it et. al.} observed that the symmetry of the magnetic degrees of freedom are broken when charge order occurs on the molecular stack \cite{Yasin12,Dressel12a}. 
Dutoit {\it et al.} confirmed these observations and pointed out that also the counter-ions $X$ are involved, in accord with previous observations \cite{Chow00,Foury10b,deSouza08,Rose12}.
Since two magnetically inequivalent sites are present below the charge-ordering temperature, the ESR spectra should show two distinct signals. 
However, exchange coupling mixes the lines, resulting in a significant increase of the linewidth.
A charge-ordered state influences the relaxation mechanism leading to a change in the anisotropy of the linewidth; the $g$-tensor remains unaffected.
In the absence of charge order, the linewidth typically follows the sinusoidal anisotropy of the $g$-value with $180 ^\circ$ symmetry.
When charge order sets in below $T_{\rm CO}$, the periodicity of the linewidth doubles, as observed when rotating the magnetic field in the $ab^{\prime}$-plane, normal to the  long axis of the molecule.
This anisotropy is enhanced with the magnetic field $H_0$ oriented in a direction between the stacking axis $a$ and the perpendicular direction $b^{\prime}$ due to anisotropic Zeeman interaction \cite{BeciniGatteschiBook} resulting  from relaxation processes between the magnetically inequivalent sites on the adjacent stacks in the charge-ordered state \cite{Yasin12}.

In the case of non centro-symmetric anions, the situation is more complex: in addition to a charge-order transition, some of the salts exhibit anion ordering at low temperatures, leading to a tetramerization of the stacks \cite{Pustogow16}. 
For \ClO{}, on the other hand, the anions order at $T_{\rm AO}=73$~K without any sign of charge-order \cite{Rosslhuber18}.
There is an ongoing discussion whether \NO{} undergoes a charge-order transition upon cooling.
DC conductivity shows a maximum at \SI{210}{K} below which the sample exhibits an insulating behavior \cite{Coulon82}.
Based on a small feature of the asymmetry of the ESR lineshape, Coulon {\it et. al.} conclude that charge order occurs around \SI{110}{K} \cite{Coulon15}.
At $T_{\rm AO}=\SI{50}{K}$ the system shows an anion-ordering phase transition leading to the formation of spin singlets. 
The exponentional decrease of the ESR intensity evidences the opening of a spin gap \cite{Moret83,Coulon15}.
Here we present results on \NO{} revealing a change in anisotropy periodicity of the ESR linewidth below around \SI{250}{K}, as previously found for other members of the TMTTF family at their charge-ordering temperature \cite{Yasin12}. 
From the similarity to the behavior in the Fabre salts with centrosymmetric anions \cite{Yasin12} we conclude a charge ordering in the title compound.

\section{Experimental details}
The measurements of the electron spin resonance (ESR) were performed on a continuous-wave X-band spectrometer (Bruker EMXplus) in a frequency range of 9 to \SI{10}{GHz}. 
The spectrometer is equipped with a goniometer.
A continuous He-gas flow cryostat (Oxford instruments, ESR 900) allows us to cool from ambient temperature to $T=\SI{4}{K}$.
The experiments were performed on single crystals of \NO{} grown by standard electrochemical procedure \cite{Liautard82a,Liautard82b}.
The specimens used for our measurements have a typical size of $2.5 \times  0.5 \times 0.1 { \rm \: mm}^3$.
The triangular anion NO$_3$ is not centro-symmetric. The TMTTF salts have a triclinic symmetry, the planar TMTTF molecules extend along the $c$-axis and are stacked in $a$-direction. 
The $g$-tensor is oriented in $a$, $b^{\prime}$, and $c^*$ direction, where $b^{\prime}$ is the projection of $b$ perpendicular to the $a$-axis, and $c^*$ normal to the $ab$-plane.

\section{Results}
\begin{figure}[b]
\centering
	\includegraphics[width=1\columnwidth]{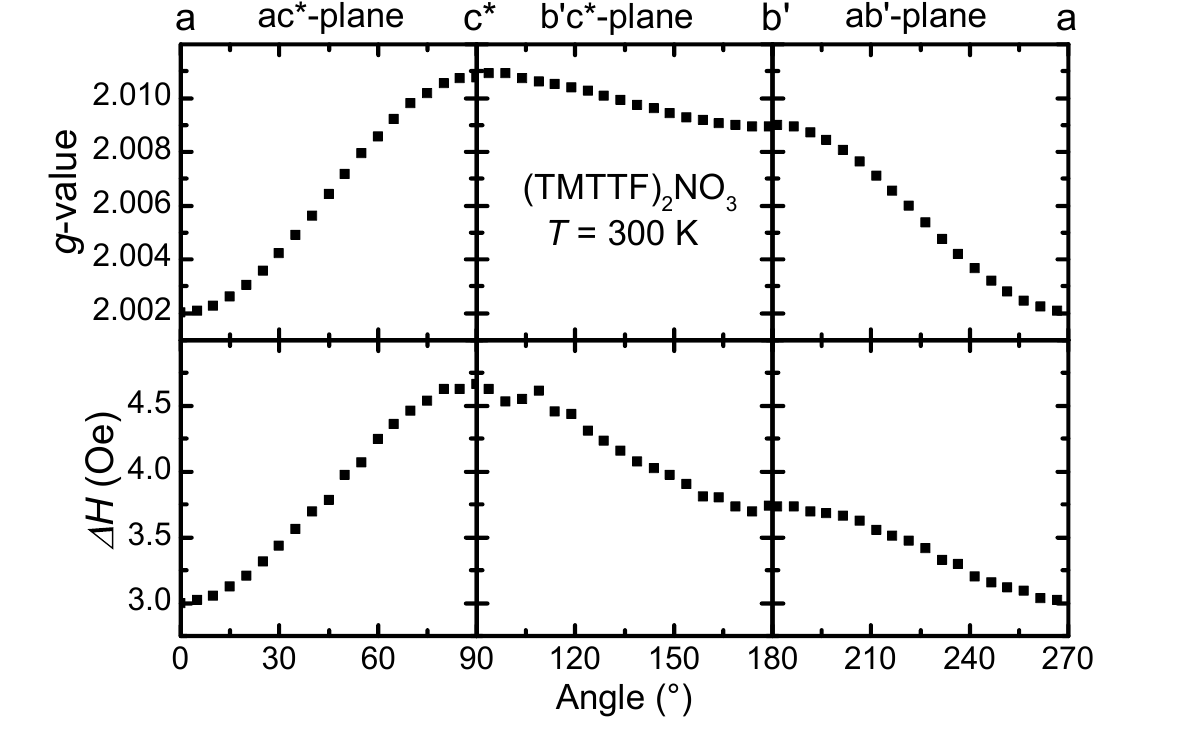}
    \caption{Anisotropy of ESR $g$-value and linewidth of \NO{}. The measurements were performed at room temperature by rotating the magnetic field $H_0$ in the $ab^{\prime}$, $ac^*$ and $b^{\prime}c^*$-planes.
    The maximum $g$-value and linewidth is found along the $c^*$-direction; the smallest  along the $a$-axis.}
  \label{fig:RT_anisotropy}
\end{figure}

The room-temperature anisotropy of $g$-value and linewidth $\Delta H$ of \NO{} is plotted in Fig.~\ref{fig:RT_anisotropy}. The maximum values of $g$ and $\Delta H$  are found with $H_0 \parallel c^*$.
Along the stacking direction $a$ the values are smallest, whereas for $H_0 \parallel b^{\prime}$ intermediate values are observed. 
Note, the linewidth always follows the behavior of the ellipsoidal $g$-tensor, i.e.\ maxima and minima alternate when rotating in steps of $90 ^\circ$.
These findings are similar to the anisotropy observed in other compounds of the \TTFX{} family.
The anomalously large rotation of the $g$-tensor upon cooling associated with the CO transition observed for \TTFX{} ($X=$PF$_6$, SbF$_6$) \cite{Furukawa09} is not present in \NO{}.
Only a minor reorientation of $\sim 3^\circ$ between room temperature and $T=\SI{4}{K}$ is observed similar to the case of (TMTTF)$_2$Br \cite{Furukawa09}.

\begin{figure}
\centering
    \includegraphics[width=1\columnwidth]{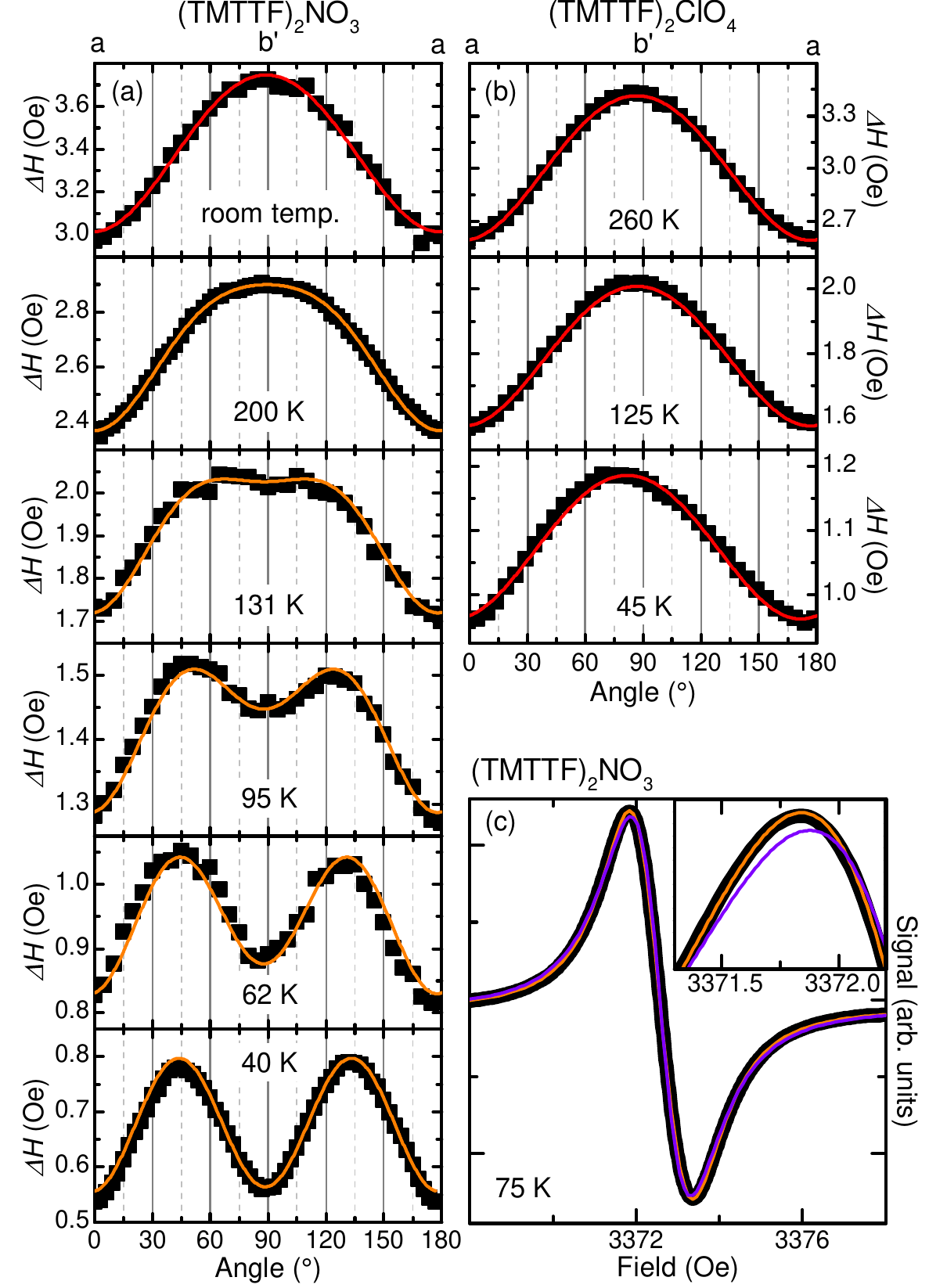}
    \caption[]{(a)~The linewidth of a \NO{} crystal measured in the $ab^{\prime}$-plane at different temperatures.  At $T=\SI{300}{K}$ the data can be fitted by Eq.~(\ref{eq:RT}), corresponding to the red line.
    For lower temperatures, $T=\SI{200}{K}$, \SI{131}{K}, \SI{95}{K}, \SI{62}{K} and  \SI{40}{K}, a sum of Eqs.~(\ref{eq:RT}) and (\ref{eq:Hmod}) have to be used; these are represented by the orange lines.
    (b)~For comparison, the angle dependence of the linewidth of a \ClO{} sample in the $ab^\prime$-plane is plotted for 
    $T=\SI{260}{K}$, \SI{125}{K} and \SI{45}{K}, respectively. 
    In this case, the signal can always be described by 
     Eq.~(\ref{eq:RT}).
    The obtained fit parameters are summarized in Tab.~\ref{AnpassparameterClO4}.
    (c)~The $T=\SI{75}{K}$ ESR signal of \NO{} in the diagonal $ab^{\prime}$-direction. The inset magnifies the maximum. The purple line was obtained by one Dysonian line, while the orange line is fitted by two Dysonian. Although the difference is small, the orange line fits better.}
  \label{fig:lowT_anisotropy}
\end{figure}

For elevated temperatures the line is slightly asymmetric due to considerable  conductivity; 
therefore a Dysonian lineshape is used for the analysis \cite{Freeman55}. 
The signal measured along the crystal axes can be well described by a single Dysonian line in the entire studied temperature-range.
In the charge-ordered  state one might expect two distinguable lines due to the magnetically inequivalent sites on the adjacent stacks. 
However, exchange interaction can mix the lines making them indistinguishable;
this results in an increased linewidth \cite{Yasin12}.
For \NO{} only slight hints of two lines are observed, as displayed in Fig.~\ref{fig:lowT_anisotropy}(c).
Using a single line, the maximum and minimum of the spectrum can be fitted almost perfectly.
We conclude that the exchange coupling dominates, as previously found for other members of the TMTTF family \cite{Yasin12}.
For all further results, we used a single Dysonian line to fit the ESR spectra.

Fig.~\ref{fig:lowT_anisotropy}(a) displays the observed $ab^{\prime}$-plane anisotropy of the linewidth for different temperatures. 
The regular ellipsoidal anisotropy of the linewidth $\Delta H(\theta)$ can be fitted by
\begin{equation}
\label{eq:RT}
\Delta H (\theta ) = \sqrt{\Delta H^2_{a}\cos^2 (\theta)+\Delta H^2_{b^{\prime}}\sin^2 (\theta)}.
\end{equation}
here $\Delta H_a$ and $\Delta H_{b^{\prime}}$ denote the linewidths for the external field $H_0$ aligned parallel to the crystallographic $a$-axis and $b^{\prime}$-axis, respectively \cite{Yasin12}.
The room-temperature behavior is well described by equation~(\ref{eq:RT}) as shown in Fig.~\ref{fig:lowT_anisotropy}(a).
Upon cooling, at around \SI{250}{K}, the  angular behavior of the linewidth  in the $ab^{\prime}$-plane starts to develop additional contributions around the diagonal directions at $45 ^\circ$ and $135 ^\circ$, and increasingly deviates from the description by equation (\ref{eq:RT});  below $T=\SI{100}{K}$ the anisotropy changes significantly.
Additional maxima develop at the positions of $45 ^\circ$ and $135 ^\circ$, exceeding the linewidth along the $b^{\prime}$ direction.
As the linewidth anisotropy starts to deviate from the behavior described by equation (\ref{eq:RT}) and indicate an onset of a doubled periodicity component (see Fig. \ref{fig:lowT_anisotropy}(a), $T=\SI{200}{K}$), an additional contribution is required to describe the angular dependence observed in \NO{}.
The doubling of the periodicity is accounted for by adding the term
\begin{equation}
\label{eq:Hmod}
\Delta H_\text{doubl.} (\theta ) = \Delta H_\text{mod}  \sin^2(2\theta)
\end{equation}
to equation~(\ref{eq:RT}). 
The amplitude of the additional contribution to the linewidth along the $45 ^\circ$ direction is given by $\Delta H_\text{mod}$.

\begin{figure}
\centering
	\includegraphics[width=1\columnwidth]{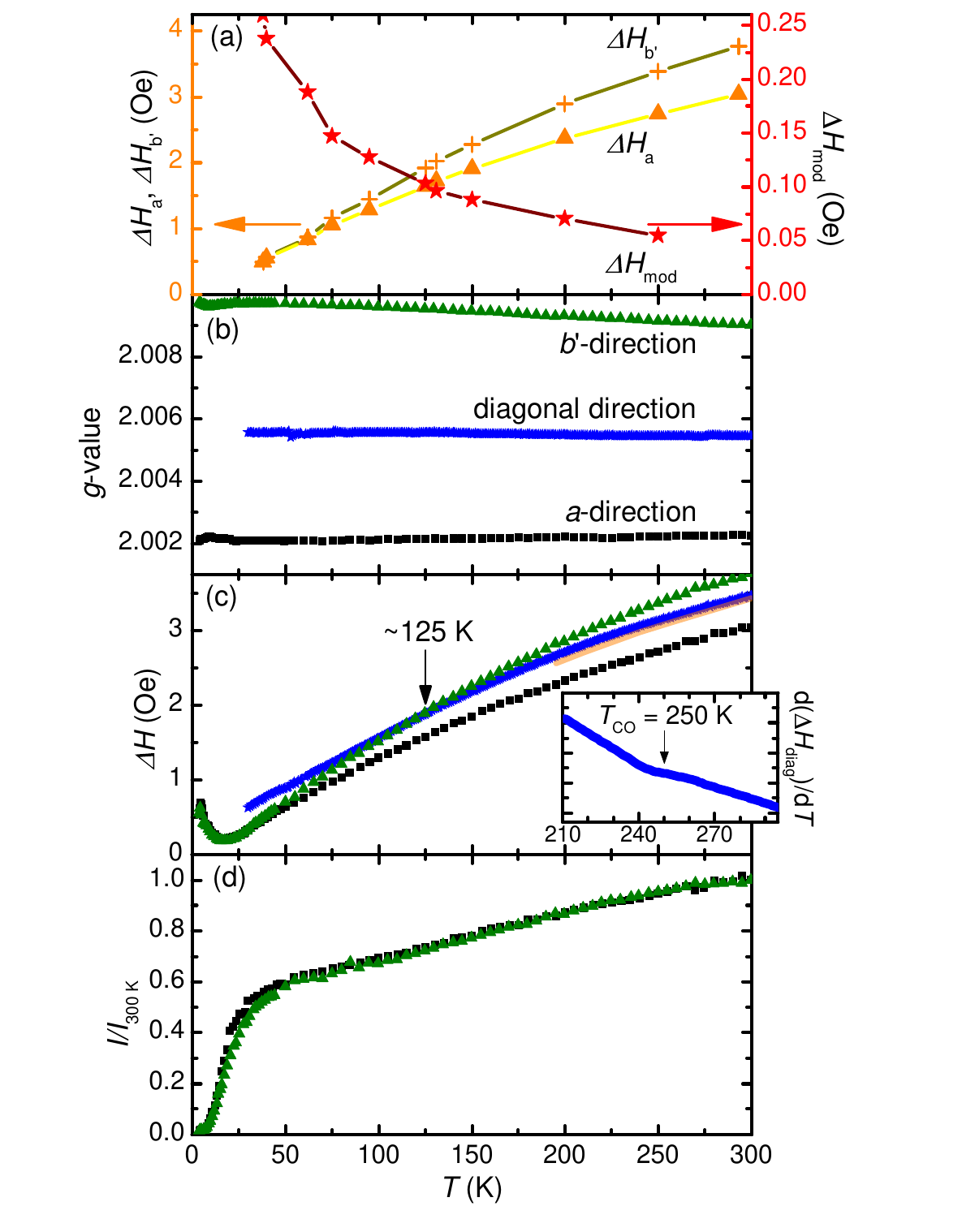}
    \caption{(a) Fit parameters of the linewidth anisotropy of the \NO{} compound at different temperatures. A sum of Eqs. (\ref{eq:RT}) and (\ref{eq:Hmod}) was fitted. Temperature dependent $g$-value (b) and linewidth $\Delta H$ (c) of \NO{} measured by X-band ESR.
    Around $T=\SI{125}{K}$, the linewidth in the diagonal direction crosses the linewidth for the $b^{\prime}$-orientation. This is taken as indication that a charge-order transition is present in \NO{}. The orange line is the estimated $\Delta H_\text{diag}$ by Eq. (\ref{eq:RT}) using the measured values $\Delta H_a$ and $\Delta H_{b^{\prime}}$.
    The inset shows the derivative of the smoothed data for the linewidth in diagonal orientation.
    (d) ESR intensity for $a$ and $b^{\prime}$-orientation normalized to the room temperature value. }
  \label{fig:dH-g-NO3-probe}
\end{figure}

Below  $T \sim \SI{250}{K}$, the angular-dependent linewidth of \NO{} was fitted by a sum of equations (\ref{eq:RT}) and (\ref{eq:Hmod}). 
Fig. \ref{fig:dH-g-NO3-probe}(a) shows the temperature dependence of the fit parameters.
The finite value of $\Delta H_\text{mod}$ suggests that a charge-order state already exists in this temperature range.

Our results resemble the behavior observed in (TMTTF)$_2$SbF$_6$, (TMTTF)$_2$AsF$_6$ and (TMTTF)$_2$PF$_6$, which undergo a charge-order transition upon cooling below $T_{\rm CO}$ \cite{Yasin12,Dressel12a}.
It is interesting to note that this doubled periodicity of the linewidth was only observed in Fabre salts with centro-symmetric anions, while in the case of linear anions, such as (TMTTF)$_2$SCN, a totally different behavior was found.
The triangular symmetry of the NO$_3^-$ anion also allows a doubling of the periodicity at lower temperatures.
From our ESR results we conclude that \NO{} exhibits a charge-order transition leading to additional relaxational processes due to anisotropic Zeeman interaction, similar to the behavior in compounds with centro-symmetric anions \cite{Yasin12}.

For comparison, we performed temperature and angular-dependent ESR measurements on \ClO{} that is known to not develop to a charge-order phase at low temperatures \cite{Coulon15}. 
As shown in Fig.~\ref{fig:lowT_anisotropy}(b), the overall behavior of the linewidth  periodicity in the $ab^{\prime}$-plane is temperature independent. 
It can be fitted using only equation~(\ref{eq:RT}).
The resulting parameters of the fit are given in Tab.~\ref{AnpassparameterClO4}.
\begin{table}
\centering
\caption{Fit parameters of the angle-dependent linewidth of \ClO{}. The measurement was performed in the $ab^\prime$-plane.
The rotation at $T=\SI{260}{K}$, \SI{125}{K} and \SI{45}{K}  can be well described by Eq.~(\ref{eq:RT}), as illustrated in Fig.~\ref{fig:lowT_anisotropy}(b).}
\begin{tabular}{l|ccc}
\ClO{}   & \SI{260}{K}  & \SI{125}{K}   & \SI{45}{K}  \\
\hline
$\Delta H_{a}$ [Oe] & $2.6\pm 0.1$  & $1.6\pm 0.1$  & $1.0\pm 0.1$  \\
$\Delta H_{b^{\prime}}$ [Oe] & $3.4\pm 0.1$  & $2.0\pm 0.1$   & $1.2\pm 0.1$  
\end{tabular}
\label{AnpassparameterClO4}
\end{table}

In Fig.~\ref{fig:dH-g-NO3-probe} the temperature dependence of the $g$-value (b) and the linewidth $\Delta H$ (c) of \NO{} is presented, as measured along the $a$ and $b^{\prime}$-direction and for $45 ^\circ$  between $a$ and $b^{\prime}$ (diagonal direction).
The $g$-value measured along $45 ^\circ$ is between the value along the $a$ and $b^{\prime}$-axes over the entire temperature range.
In contrast, the value of the linewidth in the diagonal $ab'$-direction is only halfway between the value along the $a$ and $b'$-axis for high temperatures.
However at $T^*\approx \SI{125}{K}$ the linewidth in the diagonal direction of the $ab^{\prime}$-plane becomes larger than along the $a$ and $b^{\prime}$ directions.
This confirms that in the $45 ^\circ$ direction a further contribution to the linewidth exists as already shown with the fits to the angle-dependent measurements. 
However, already at a temperature of around $T=\SI{250}{K}$ the linewidth of the diagonal-direction $\Delta H_\text{diag}$ is larger than the value estimated by equation (\ref{eq:RT}) using the measured values $\Delta H_{a}$ and $\Delta H_{b^{\prime}}$, as indicated by the solid orange line in \ref{fig:dH-g-NO3-probe}(c).

The temperature dependence of the linewidth $\Delta H(T)$ shows no distinct features along any crystallographic axes in contrast to the behavior observed for (TMTTF)$_2$SbF$_6$ and (TMTTF)$_2$AsF$_6$, where a clear kink has been found at the charge order temperatures \cite{Yasin12}.
The behavior resembles (TMTTF)$_2$PF$_6$: there at the charge-order transition at $T$ = \SI{67}{K} no kink becomes obvious.
Only for the diagonal orientation a change in the slope of the derivative (d($\Delta H_\text{diag}$)/d$T$) is identified around $T_\text{CO}=\SI{250\pm 10}{K}$, as indicated in the inset of Fig.~\ref{fig:dH-g-NO3-probe}(c).
Hence, for our ESR data we suggest that \NO{} is charge ordered below $T_\text{CO} \approx \SI{250}{K}$.

Fig.~\ref{fig:dH-g-NO3-probe} (d) shows the temperature dependence of the ESR intensity. 
The overall behaviour resembles the spin susceptibility of other \TTFX{} compounds \cite{Dumm00a}.
The intensity monotonically decreases upon cooling from room temperature to \SI{4}{K}.
Below the anion-ordering transition at around \SI{50}{K} the chain dimerizes accompanied by the formation of spin singlets, leading to the observed exponential decay of the spin susceptibility \cite{Moret83,Coulon15}.

\section{Conclusion}
In summary we have performed angular and temperature dependent X-band ESR experiments on \NO{} and \ClO{} single crystals. 
The temperature evolution of the linewidth anisotropy in the $ab^\prime$-plane for \NO{} gives microscopic evidence for the stabilization of a charge ordered phase below $T_\text{CO}=\SI{250\pm 10}{K}$ for this compound.

\begin{acknowledgements}
We are very grateful to Dr. Yohei Saito and Dr. Roland R\"{o}sslhuber for fruitful discussions.
\end{acknowledgements}

\bibliographystyle{spphys}       
\bibliography{literatur}

\end{document}